# Energy Scale of the Self-Energy in High-$T_C$ Superconductors

E. G. Maksimov

*P. N. Lebedev Physical Institute, Leninskii prosp. 53, 119991 Moscow, Russia*

maksimov@lpi.ru

The contradictions between some recent statements about the origin and the doping dependence of a sharp peak in the electron self-energy obtained from optics and ARPES are discussed. It is shown that the conclusion of Hwang *et al.* about disappearance of a mass renormalization in the highly overdoped regime is unfounded.

Recently a few contradicting statements have been made about the origin and the doping dependence of a sharp peak in the electron self-energy from optics and ARPES experiments [1–3]. The strong electron mass-renormalization has been observed a few years ago by ARPES (see corresponding refs. in [1–3]). It was clear that this renormalization is due to the interaction of electrons with some Bose-type excitations which energies are considerably small ( ~ 0.1 eV). These bosons have been related to a magnetic resonance and phonons (see again refs. in [1–3]).

In a recent study of the optical conductivity of the Bi-2212 system Hwang *et al*. [1] have confirmed the mass-renormalization changes observed in ARPES but claim that such changes are no longer observed in the highly overdoped regime at the doping level of 0.23. Taking into account that the transition temperature at this doping level is still considerably high ($T_c$ = 55 K ) Hwang et al.[1] concluded that both the magnetic resonance and phonons can be ruled out as the principal cause of high- $T_c$ superconductivity.

In a comment on this study, Cuk *et al*. [2] emphasized that the conclusion of Hwang *et al*. is unfounded for some reasons. In the response to this Cuk *et al*. comment, the authors of the work [3] have answered only on one reason mentioned by Cuk *et al*. It was related to the exact value of $T_c$ of an overdoped sample where the mass-renormalization has been observed by Cuk *et al.* in the ARPES measurements. It is difficult for me to participate in the discussion about the exact value of $T_c$ in some concrete sample then I would like to emphasis here that there are more serious reasons to consider the conclusion of Hwang *et al.* as unfounded.

It was mentioned by Cuk *et al*. [2] that optics measures a momentum average and therefore is not a very sensitive probe when the signal is strongly momentum dependent as it is for the electron self-energy in high- $T_c$ superconductors. Moreover, Hwang *et al*. use the generalized Drude formula to describe the optical conductivity both for the normal and superconducting states. As it was shown in our review [4] the expression for the conductivity in the generalized Drude form can be derived exactly only for the normal state of nearly isotropic systems. For this case there is the exact expression for the optical self-energy in the terms of the spectral function of a corresponding electron-boson interaction [4]. Using this expression, one can make the definite conclusions about the interrelation between the temperature and frequency dependence of the optical self-energy and the behavior of the boson spectral function. Moreover, as was shown many years ago by Allen [5], in this case the second derivative of the optical relaxation rate for the normal state at $T = 0$ will give the corresponding spectral function. It will be very interesting to do such measurements for very high doping level where $T_c = 0$. I would like also to emphasis here that the temperature and frequency dependence of the optical self-energy is very different from that of one-particle self-energy. For the example, the one-particle relaxation rate become to be constant at the energies coinciding with the maximal boson energy but the optical relaxation rate continues to increase up to energies ten times larger than the boson energy.

The generalized Drude formula can be used for anisotropic systems only as some type phenomenological approach and no any definite relations exist between optical self-energy and the corresponding spectral functions. Everybody should be care now to make any definite conclusions about electron-boson interaction and its changes if only the changes of such phenomenological self-energy are known. The generalized Drude formula is certainly inapplicable for a superconducting state because the conductivity of superconducting state depends both on the normal self-energy and the superconducting order parameter as well as on corresponding coherence factors

Now let me return to the discussion of Ref. [1]. These authors separate the real part of the self-energy on the sharp feature and a broad background. As can be seen from Figs. 1 and 2, the sharp feature exists only in the superconducting state and it is nothing more than the artifact of their representation of the optical self-energy for a superconducting state in the form of the generalized Drude formula. The amplitude of this artificial resonance peak will depend not only on the spectral function of the boson excitations but also on superconducting properties of the system. That are a gap value and its symmetry. Something similar have been observed in the paper [1] and it is shown on the Fig. 3. The real part of their optical self-energy in the normal state has a broad maximum at the energies between 1000 and 2000 cm$^{-1}$. That is the feature which has been named in the paper [1] as the background and its form does not contradict to the supposition about the interaction of the electrons with phonons. The authors [1] have mentioned in their article correctly that the slope of the self-energy at small frequencies is closely related to the constant of coupling of electrons with the corresponding bosons. This slope, as it can be easily seen from Fig.1 of this work, does not equal zero at all studied doping levels at least in the normal state where it make some real sense. The existence of the real background in the one-particle self-energy besides of the electron-phonon interaction has been observed recently [6] by ARPES measurements in the LSCO systems.

In conclusion, the detail discussion of the results obtained in the paper [1] is not the subject of this comment but I believe that the presented arguments have demonstrated that the conclusion of Hwang *et al*. [1] is certainly unfounded.